# Coherent broadband mid-infrared supercontinuum generation in $As_2Se_3$ photonic crystal fiber

**Wu Yuan**[1,*], **Songlin Zhuang**[1]

[1]*Shanghai Key Lab of Contemporary Optical System, Optical Electronic Information and Computer Engineering College, University of Shanghai for Science and Technology, Shanghai 200093 China*
*\* scottwyuan@gmail.com*

**Abstract:**

The generation of fully coherent broadband mid-infrared (MIR) supercontinuum (SC) from ~2.3 μm to ~8.3 μm is demonstrated by using a 4.1 um pump and an $As_2Se_3$ photonic crystal fiber (PCF). By introducing the random quantum noise and the power instability on the input pulse and by numerically implementing the Young's double slits experiment, we examine the coherence properties across the SC spectrum. It is found that the coherence of this MIR SC source depends strongly on the input pulse duration, the peak power, the power stability, and the zero-dispersion wavelength (ZDW) of the $As_2Se_3$ PCF. The optimal conditions for the MIR SC with a maximal coherent bandwidth are identified.

## 1. Introduction

There is currently a world-wide push towards applications of light in the MIR wavelength regime [1-4]. Light sources in this wavelength range have wide applications in medical diagnostics, green chemistry, and spectroscopy, as well as in the food, pharmaceutical, and defense industry such as infrared counter-measures. SC generation in optical fiber is a reliable and economical way to achieve the dramatic spectral broadening in the MIR regime [3,4]. This technique has been widely demonstrated and well understood in all pump regimes ranging from continuous wave, nanosecond, picosecond to ultrashort femtosecond [5]. Commercial SC sources are now available using primarily lasers at 1064 nm and can provide a bandwidth from 400–2400 nm.

With conventional silica glass fibers wavelengths longer than 2.4 μm cannot be accessed simply because of their near-infrared (NIR) loss edge. The development of so-called soft glass fibers, such as fluoride, chalcogenide, and tellurite fibers, have attracted more and more research interests, because they guide well in the MIR regime and can have a factor of ~1000 higher nonlinearities than silica [6-13]. Among them, $As_2Se_3$ glass has the highest nonlinear index [7,8] and the widest MIR tramission window which cuts off at about 10 μm at the long-wavelength side [8,13]. It potentially allows the generation of significant SC radiation beyond 6 μm, which is highly desirable for variant advanced MIR spectrascopic applications.

However, $As_2Se_3$ glass has a larger refractive index than both fluriode and tellurite glass [8], which results in a longer ZDW at about 5 μm. It is difficult to find a pump source to work around the ZDW of $As_2Se_3$. As a reliable pump source in 1.8 to 2.1 μm wavelength regime, mode-locked thulium-doped fiber laser has attracted increasing interests [10-12]. The subharmonic generation source of the mode-locked thulium-doped fiber laser from a degenerated optical parametric oscillator (OPO) is a highly potential pump for generating MIR wavelength beyond 6um with the $As_2Se_3$ fiber. This new pump scheme is very compact and useful. It has been successfully demonstrated by pumping a tapered chalcogenide fiber with a subharmonic generation source of the mode-locked erbium-doped fiber laser to achieve a broadband MIR SC from 2–4.6 μm [14]. By using the subharmonic generation source of the mode-locked thulium-doped fiber laser at ~4.1 μm [15], while employing the PCF as a valid method to tailor the dispersion of $As_2Se_3$ fiber and move the ZDW down to a shorter wavelength, the generation of a hyper-broadband from ~2 μm to ~10 μm with a remarkably spectral flatness has recently been reported by us [3].

Despite the broad spectral bandwidth, the sensitivity of corresponding coherence properties of this highly potential MIR SC source to the input pulse noises has not yet been the subject of a detailed study. This aspect is of particular importance for the coherent spectral broadening of a MIR frequency comb to a multi-octave-spanning one with the stablized carrier-envelope offset (CEO) frequency and repetition frequency, which are desirable for the applications such as the precision MIR optical frequency metrology and spectroscopy [1, 14]. Previous works have examined the amplitude and phase noises on the SC broadened frequency comb resulting from both quantum noise and technical noise (i.e. peak power fluctuations) on the input pulse in the NIR regime [16-18]. It is found that both the amplitude and phase noises are related to the optical coherence degradation across the comb, which manifests in the phase jitter, amplitude jitter, timing jitter, and frequency jitter [17]. The measurement of the wavelength depedent coherence provides a straightforward method to understand the noise amplification mechanisms in the SC generation.

A well-established technique by using a Young's double slits experiment can be readily carried out to measure the polychromatic far-field interference pattern and resolve the fringe visibility at each wavelength in the spectrum [19]. Numerically this measure of coherence is implemented by carrying out the simulations in the presence of input pulse noises, i.e., quantum noise and pulse power fluctuations, and using the ensemble average over a large number of independent SC pairs with which each pair resembles the two independent sources emitting from the double slits [18]. In this paper, we examine numerically the coherence properties of the MIR SC generated in $As_2Se_3$ PCF. The coherence is shown to depend strongly on the input pulse's duration, the peak power, the power stability, and the ZDW of the $As_2Se_3$ PCF. We identify the optimal conditions for the generation of a fully coherent MIR SC with the maximal bandwidth.

## 2. Mid-infrared SC generation model

The generalized nonlinear Schrödinger equation (GNLSE) is employed to study the pump pulse propagation and the MIR SC generation. We consider the material loss, higher order dispersion, stimulated Raman scattering, and frequency dependence of the nonlinear response in our simulations. A change in variables is used to shift into so-called interaction picture in order to avoid the stiff dispersive part of the GNLSE and obtain the ordinary differential envelope equation [3,20,21]:

$$\frac{\partial \widetilde{A'}}{\partial Z} = i\bar{\gamma}(\omega)\exp{(-i\hat{L}(\omega)Z)} \times \mathcal{F}\left\{A(Z,T)\int_{-\infty}^{\infty}R(T')|A(Z,T-T')|^2 dT'\right\} \quad [1]$$

where the nonlinear response $\bar{\gamma}(\omega)$ is given by

$$\bar{\gamma}(\omega) = \frac{n_2 n_0 \omega}{cn_{eff}(\omega)A_{eff}^{1/4}(\omega)}$$

And

$$A(Z,T) = \mathcal{F}^{-1}\left\{\frac{\tilde{A}(Z,\omega)}{A_{eff}^{1/4}(\omega)}\right\}$$

Here $\mathcal{F}$ denotes Fourier transformation. The change in variables is given by $\widetilde{A'}(Z,\omega) = \tilde{A}(Z,\omega)\exp{(-i\hat{L}(\omega)Z)}$, in which the linear operator $\hat{L}(\omega) = \beta(\omega) - \beta(\omega_0) - \beta_1(\omega_0)[\omega - \omega_0] + i\alpha(\omega)/2$.

We derive $h_R(t)$ from the Raman gain spectrum measured by Naval Research Laboratory [22] in order to determine the Raman response function $R(t) = (1-f_R)\delta(t) + f_R h_R(t)$. It is found that the measured gain spectrum can be modeled by a two decaying harmonic oscillators function with the fractional contribution of $f_a$ and $f_b$ respectively [3]:

$$h_R(t) = f_a\tau_1(\tau_1^{-2} + \tau_2^{-2})\exp{(-\tau/\tau_2)}\sin(\tau/\tau_1) + f_b\tau_3(\tau_3^{-2} + \tau_4^{-2})\exp{(-\tau/\tau_4)}\sin(\tau/\tau_3) \quad [2]$$

By choosing $f_a = 0.7, f_b = 0.3, \tau_1 = 23 \times 10^{-15}, \tau_2 = 230 \times 10^{-15}, \tau_3 = 20.5 \times 10^{-15}, \tau_4 = 260 \times 10^{-15}$, our model fits the experimental results of ref. 19 very well with the main peak of the Raman gain located at $\Omega_R$ ~6.9 THz (~230 cm$^{-1}$) [22].

The nonlinear index $n_2$ of chalcogenides as estimated by using the explicit Kramers-Kronig transformation equation is found to be $n_2 = 5.74 \times 10^{-18}$ m$^2$/W at wavelength of 4.1 μm [3, 23,24]. This value is approximately 220 times the nonlinear index of silica at wavelength 1.5 μm, which is ~ 2.6 $\times 10^{-20}$ m$^2$/W. We use $f_R = 0.1$ as reported by Slusher et al [22]. The two photon absorption coefficient is neglected because the interested spectral region is far away from the main peak of two photon absorption at 1.39 μm [22].

For the refractive index of $As_2Se_3$, the experimentally derived Sellmeier equation proposed by Thompson of Amorphous Material Inc. at 2008 [25] is employed in our calculations thereafter:

$$n(\lambda) = \left[1 + \lambda^2[A_0^2/(\lambda^2 - A_1^2) + A_2^2/(\lambda^2 - 19^2) + A_3^2/(\lambda^2 - 4 \times A_1^2)]\right]^{0.5} \quad [3]$$

where $\lambda$ is the wavelength, $A_0, A_1, A_2, A_3$ are Sellmeier coefficients and $A_0=2.234921, A_1=0.24164, A_2=0.347441, A_3=1.308575$, respectively.

## 3. Fiber design

$As_2Se_3$ PCF with the conventional five-ring triangular structure is considered because of its ease of fabriction by utilizing standard extrusion and stacking based methods. The effective index of the fundamental mode and its corresponding confinement loss of $As_2Se_3$ PCF at the wavelength of interest are calculated for different pitches ($\Lambda$= 3, 4, 5, 6 μm) by using the finite-element method based full-vector solver COMSOL. The relative hole size (d/ $\Lambda$ = 0.4) is kept constant to assure that the fibers are endlessly single-mode [26].

The resulting dispersion and confinement losses for the $As_2Se_3$ PCF of different pitches are shown in Fig 1. In order to achieve the maixaml SC bandwidth limited only by the intrinsic material absorption at < ~2 μm and > ~9 μm [8,22], the $As_2Se_3$ PCF shall has a pitch of more than 3 μm to avoid the negative impacts of the confiment loss, in particular, at long wavelengths above ~7 μm. For the pitches ≥4 μm, it is noted that a low and flat dispersion profile with an absolute dispersion within 30 ps/(nm.km) can be achieved for the spectral range from ~3 μm to ~10 μm, which is important for the maximization of the SC generation in fiber as it can decrease the temperal walk-off effect during the spectral broadening process [5].

As we have demonstrated before, it turns out that the realization of the hyper-broadband MIR SC generation with a bandwidth more than 7 μm and a high spectral flatness occurs when pumping a 4.1 μm sub-picosecond pulse near to the first ZDW, such as when the pitch is 4 μm or 5 μm, after the propagation of only a few cm fiber length [3]. Therefore, in order to achieve a MIR SC generation with the maximal coherent bandwidth, the $As_2Se_3$ PCFs with 4 μm and 5 μm pitches are investigated in our following numerical studies. Note that the pumping wavelength of 4.1 μm is in the anomalous dispersion region for $\Lambda$ = 4 μm, whereas the pulse experiences normal dispersion when $\Lambda$ = 5 μm.

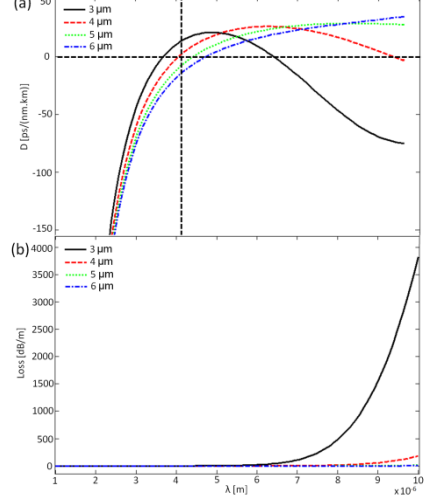

Fig.1. Dispersion profiles (a) and the confinement loss (b) of the $As_2Se_3$ PCFs with different pitches and a constant hole to pitch ratio of 0.4, black solid line: $\Lambda$ = 3 μm, red dashed line: $\Lambda$ = 4 μm, green dotted line: $\Lambda$ = 5 μm, blue dash-dotted line: $\Lambda$ = 6 μm. The vertical black dashed line in (a) indicates the pump wavelength of 4.1 μm.

## 4. Results and discussions

Simulations are first carried out with the 4.1 μm hyperbolic secant input pulse of a 5 kw peak power and a 500 fs pulse duration (FWHM) which is injected in the anomalous dispersion regime of a $As_2Se_3$ PCF with a 4 μm pitch. Noise is included by adding to the input pulse one random phase photon per mode in the frequency domain. Figure 2 shows the output spectra and temporal intensity profiles at several propagation distances obtained for a SC pair, i.e., two independent simulations performed under identical conditions except for the random quantum noise on the input pulse. The ZDW is also illustrated in the spectra plots. As we observe from Fig. 2 the symmetric sideband structures form after only a few millimeters. This suggests that, when pumping in the anomalous regime, the initial spectral broadening is due to the four-wave mixing (FWM). With further propagation, nonlinear and dispersive interactions lead to the rapid temporal oscillations owing to the ultrafast modulational instability (MI) effects [16, 17]. The MI initiates the soliton generation and soliton fission processes. Combining with the Raman self-frequency shift, a dramatic spectral broadening and the generation of a train of distinct Raman solitons are resulted.

Comparsion of the SC pair in Fig. 2 shows that the MIR SC generation is affected significantly on both the temporal and spectral intensity profiles by the input pulse noise from run to run, in particular, at the long wavelength side and the long propagation distance where the MI takes effect. These observations agree with previous studies of SC generation in conventional fibers, which have shown that broadband spectral generation seeded from MI is highly sensitive to input pulse noise [5]. Significant variations of temporal and spectral characteristics are associated with coherence degradation caused by severe fluctuations in the spectral phase and amplitude at each wavelength.

To study the coherence degradation we numerically implement the Young's double slits experiment and examine wavelength dependence of the fringe visibility across the SC spectrum. To achieve this we calculate, for each set of simulational parameters, 100 independently generated SC pairs, i.e., $E_{i,1}(\lambda, t)$ and $E_{i,2}(\lambda, t)$, obtained from input pulses with different random quantum noise. Each SC pair represents two independent sources out from double slits with a propagation time of t, and the same propagation time t would correspond to measuring the fringe visibility at the center of the fringe pattern in a Young's doule slits experiment. Such coherence property as quantified with the degree of coherence across the SC spectrum can be calculated as [18]:

$$|g_{12}(\lambda)| = \left|\frac{\langle E_1^*(\lambda,t)E_2(\lambda,t)\rangle}{[\langle|E_1(\lambda,t)|^2\rangle\langle|E_2(\lambda,t)|^2\rangle]^{1/2}}\right| \qquad [4]$$

The angle brackets denote an ensemble average and the application of the ensemble average over a large number of independent SC pairs is to represent a signal acquisition process with which a series of polychromatic far-field interference pattern are measured and averaged. Also note that the ~270 THz frequency resolution fixed by the simulations will yield an effective average coherence over many longitudinal modes for SC generated from typical 80-MHz repetition rate sources.

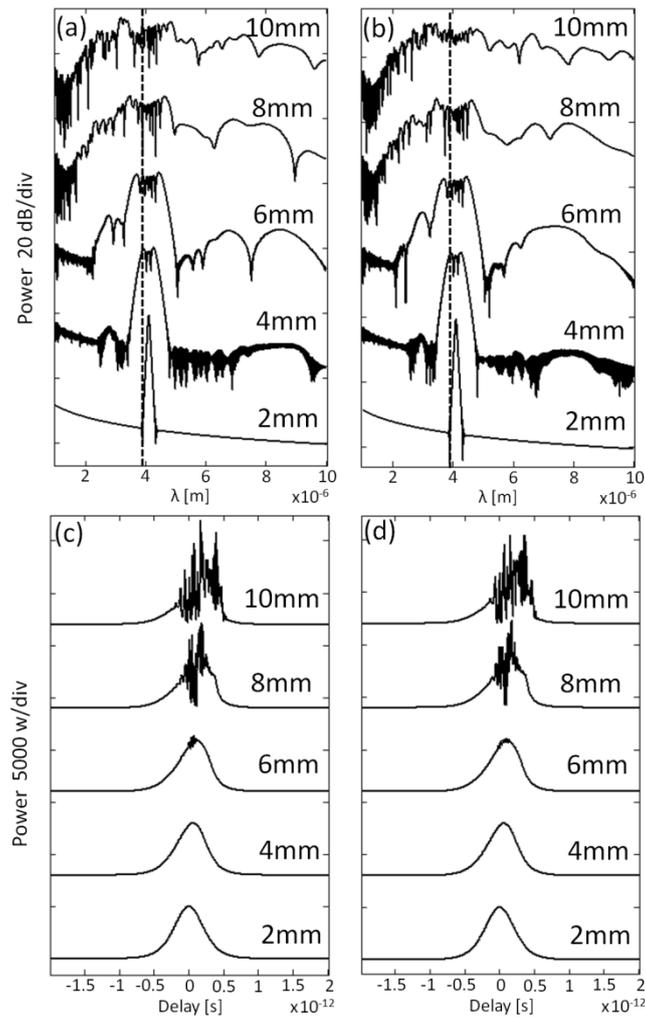

Fig. 2. Spectral (a,b) and temporal (c,d) intensity profiles of two independent simulations (a-c, b-d) with a 4.1 μm pump of a 5 kw peak power and a 500 fs pulse duration. The dotted line shows the ZDW of the $As_2Se_3$ PCF with a 4 μm pitch.

The correlation of the decoherence with the spectral broadening process is studied when pumping in both normal and anomalous regimes. As shown in Fig. 3, we plot the coherence and spectral evolutions along a 6 cm long $As_2Se_3$ PCF with differenct pitches. For both pitches, during the initial stage of SC generation, the input pulses undergo a dramatic spectral broadening, with $g_{12}(\lambda) = 1$ over majority of the SC spectrum. When pumping in the anomalous regime for $\Lambda = 4$ μm, the initial spectral broadening is due to a coherent process of FWM, while a coherent process of higher-order dispersion phase-matched FWM dominates in the initial SC generation for $\Lambda = 5$ μm. With further propagation, a significant coherence degradation with a reduced $g_{12}$ is observed for both cases. Although further propagation does lead to more spectral broadening which eventually spans from ~2 μm to ~10 μm, the coherence degradation increases dramatically to $g_{12} \ll 1$ over most of the SC spectrum at z=6 cm.

It is well understood that MI dominates thus coherent degrades for the anomalous dispersion regime pump. For pumping in the normal regime, the additional broadening is mainly due to the coherent process of self-phase modulation (SPM). However, when the SC spectrum extending into the anomalous dispersion regime, the interactions between anomalous dispersion and nonlinearity would inevitably initial the MI which degrades the coherence consequently. As it is noted from Fig. 3, although the competition between the coherent and the incoherent processes exists, pumping in the normal dispersion regime leads to a broader coherent SC generation with a 20 dB bandwidth from ~2.5 μm to ~5.5 μm, which is in contrast to a coherent spectrum from ~3.5 μm to ~5 μm for the anomalous regime pump.

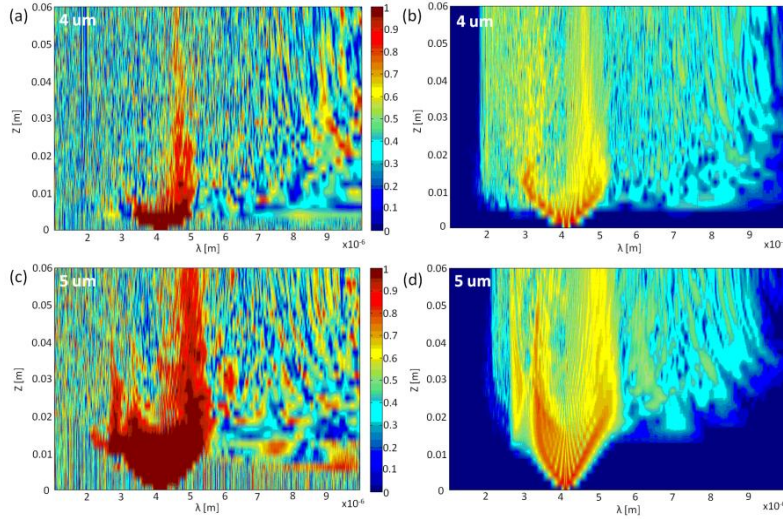

Fig. 3. The coherence (a, c) and the spectral (b, d) evolution profiles along a 6 cm long $As_2Se_3$ PCF with a 4.1 μm pump of a 5 kw peak power and a 500 fs pulse. The top curves (a, b) show for pumping in the anomalous dispersion regime with a pitch of 4 μm, the bottom curves (c, d) show for pumping in the normal dispersion regime with a 5 μm pitch, both calculated from an ensemble average.

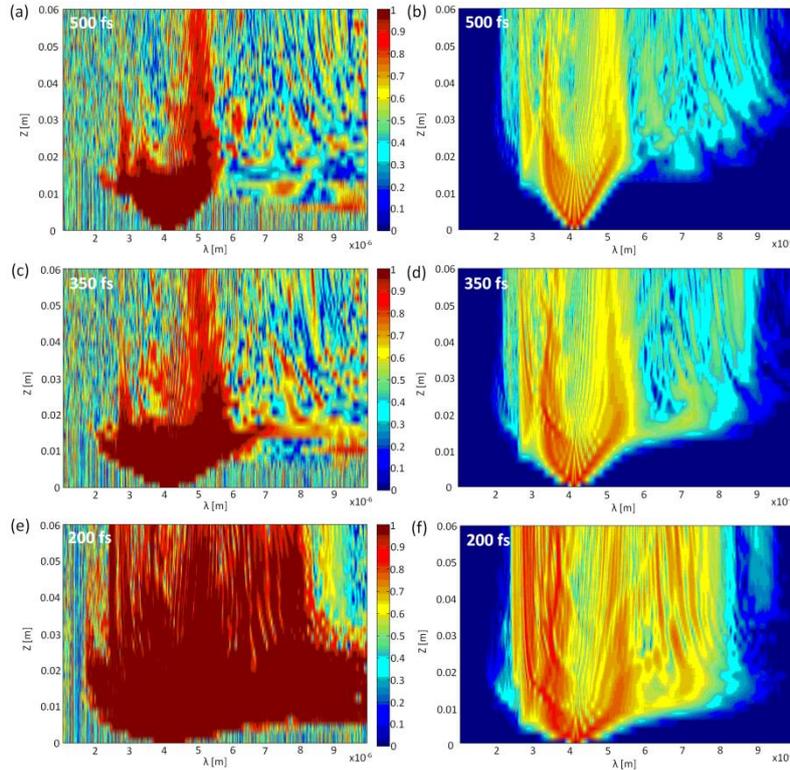

Fig. 4. The left-hand side plots (a, c, e) show the coherence evolution profiles and the right-hand side plots (b, d, f) show the spectral evolution profiles, all along a 6 cm long $As_2Se_3$ PCF with a pitch of 5 μm. All simulations are carried out with a 4.1 μm pump, a 5 kw peak power, and the pulse durations of 500 fs (a, b), 350 fs (c, d), and 200 fs (e, f), respectively.

For frequency metrology it is clear that SC must be generated under conditions in which coherence degradation is minimized, as well as the spectrum generated is as wide as possible. The observation of Fig. 3 suggests that, in order to achieve a broad coherent SC generation, the coherence processes such as SPM shall be enhanced. Therefore, using a shorter input pulses would be prefered in our conditions for which coherent SPM process plays a more significant role in the spectral broadening and the effects of MI are reduced. Figure 4 presents the confirmation of the influence of pulse durations on the coherence degradation, showing simulation results for different pulse durations in a $As_2Se_3$ PCF with a pitch of 5 μm. When pumping in the normal-dispersion regime, it can be seen that, with a short pulse duration, the SPM process can greatly override the influence of the MI effects for a longer propagation distance than that in the case of 500 fs. In particular, for a pulse of 200 fs, a coherent spectrum from ~2.8 μm to ~7.8 μm at 20 dB level can be readily achieved at z=1.8 cm, as illustrated in Fig. 4 (e,f) and Fig. 5 (c,d).

Significantly, although increasing the pitch to more than 5 μm can shift the pump wavelength further into the normal dispersion regime, and therefore further enhance the coherent SPM process, our simulations show that the generated coherent SC spectrum is significantly narrower than that obtained with pumping near the ZDW as demonstrated. To further optimize the coherent bandwidth, we investigate how the peak power of the 4.1 μm pump laser with 200 fs pulse duration would affect the coherent SC generation in a $As_2Se_3$ PCF with a pitch of 5 μm.

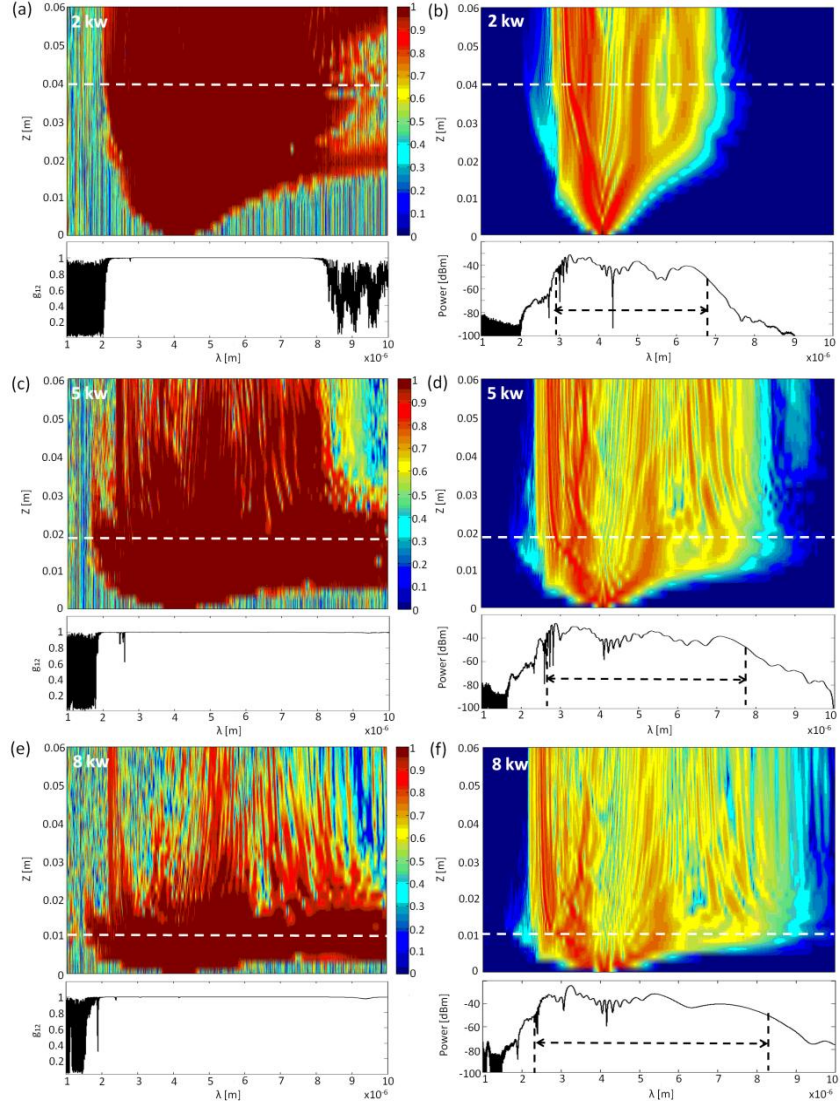

Fig. 5. The coherence (a, c, e) and the spectral (b, d, f) evolution profiles along a 6 cm PCF with a 5 μm pitch for a 4.1 μm pump with a pulse duration of 200 fs and a peak power of 2 kw (a, b), 5 kw (c, d), and 8 kw (e, f), respectively. The coherence curves and the SC spectra at the maximal coherent length for the respective peak power are showed under each plot, the maximal coherent length is indicated with white dashed lines, and the black dahsed-line arrows in each SC spectrum show the 20 dB bandwidth.

As illustrated in Fig. 5, the simulations suggest that the increase of peak power would facilitate the influence of MI effects and thus decreases the characteristic length scale for the coherent SC generation. Meanwhile, the increase of peak power accelerates the spectral broadening process, which leads to a hyper-broadband coherent SC spectrum with a 20 dB bandwidth from ~2.3 μm to ~8.3 μm at z=1 cm for 8 kw. One the other hand, the decrease of peak power would increase the coherent length, but decreases the coherent SC bandwidth to ~2.9 μm to ~6.8 μm at z=4 cm for 2 kw. This is expected since, as a dominative process at the SC generation in the femtosecond regime and with the normal dispersion regime pump, the efficiency of the FWM and SPM, i.e. its gain and bandwidth, depends on nonlinear index $n_2$, effective mode area $A_{eff}(\omega)$, and the peak power [27].

When $n_2$ and $A_{eff}(\omega)$ are constant, the low peak power results in a low FWM gain and a low efficient FWM process. This lays down the fundation for the following low efficient SPM process in the spectral broadening, although in this case the coherence processes take the competitive edge over the incoherent MI effects. In contrast, a high peak power leads to a high gain for both the FWM and SPM processes, while results in the earlier occurrence of the MI effects and coherence degradation. This is because the MI gain depends exponentially on the peak pulse power and the propagation length [5, 16].

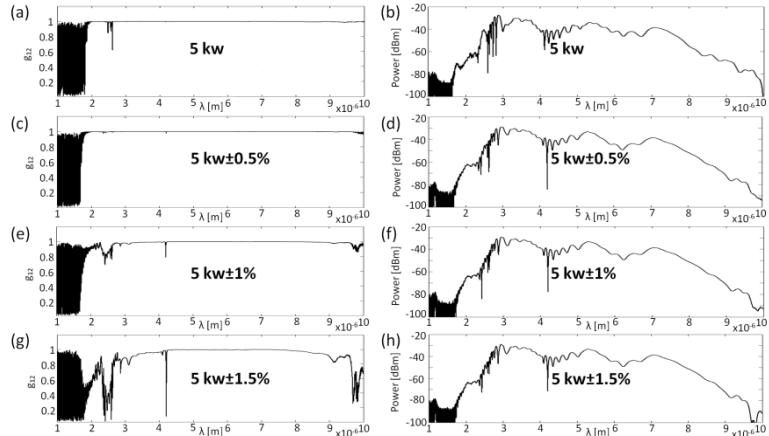

Fig. 6. The coherence profiles and corresponding SC spectra at length of 1.8 cm for different pulse to pulse power stability of (a,b) ±0%,(c,d) ±0.5%,(e,f) ±1%,(g,h) ±1.5%. A 4.1 μm pump with a pulse duration of 200 fs and a peak power of 5 kw is used.

Moreover, as shown in Fig. 6, the simulations illustrate that a peak power instability of more than ±0.5% on the pump pulses can also induce significant variations of spectral characteristics and degrades the coherence. This coherence degradation starts from two spectral edges, where the generated SC has a higher sensitivity of the amplitude noise to the input pulse power fluctuations [28]. This amplitude noise, which is always manifested in the low-frequency componet of the amplitude noise spectrum of generated SC [16,17], is directly associated with the increased coherence noise and can be readily reduced by using stabilized pump lasers with pulse to pulse power stability of less than ±0.5%. In contrast, the coherence degradation resulting from the random quantum noise on the input pulse is intrinsic and set the fundamental coherence noise limitations to the MIR SC generation. Furthermore, similar coherence properties are found in our simulations by setting the Raman gain to zero, implying that the primary cause of phase and amplitude noise, thus coherence degradation, is the sensitivity of MI to input pulse noise and not stimulated Raman scattering in our conditions [16,17,18].

## 5. Conclusions

For frequency metrology the MIR SC generation with a wide coherent bandwidth is highly desirable. Although the coherence degradation caused by the input pulse noise, in particular quantum noise, cannot be totally removed, the simulational results imply that the coherent SC generation in MIR regime can be achieved by enhancing the coherent nonlinear processes and reducing the MI effects. The decoherence induced by the power instability can also be reduced remarkable by using a stablized pump laser through feedback. With fiber parameters used in our simulations, the results suggest that these optimal conditions would require pulses of ≤ 200 fs duration, a pump at the normal dispersion regime and near the ZDW, and a peak power of ≥ 5 kw with the power stability of less than ±0.5%. With these conditions, the generation of a hyper-broadband coherent MIR SC from ~2.3 μm to ~8.3 μm can be achieved with an $As_2Se_3$ photonic crystal fiber and a subharmonic generation source of the mode-locked thulium-doped fiber laser at ~4.1 μm. This scheme is a feasible solution to achieve a coherent spectral broadening for the MIR frequency comb with a multi-octave-spanning and stable spectral phase at each wavelength. The degree of coherence provides a convenient means to quantify the coherence properties of SCs, and the experimental implementation would be readily carried out.